\documentclass[twocolumn,aps]{revtex4}
\usepackage{amsmath}
\usepackage{amssymb}
\usepackage{graphicx}
\usepackage{dcolumn}
\usepackage{bm}
\usepackage{siunitx}
\usepackage{xcolor}
\usepackage{epstopdf}
\usepackage{longtable}

\begin{document}

\preprint{Phys.Rev.B }

\title{Viscous Electron Flow and Nonlinear Magnetotransport in 2D Channels}
\author{A. D. Levin,$^1$ G. M. Gusev,$^1$  and A. K. Bakarov$^{2}$}

\affiliation{$^1$Instituto de F\'{\i}sica da Universidade de S\~ao
Paulo, 135960-170, S\~ao Paulo, SP, Brazil}
\affiliation{$^2$Institute of Semiconductor Physics, Novosibirsk
630090, Russia}

\date{\today}
\begin{abstract}
We examine nonlinear transport in a viscous two-dimensional electron fluid within narrow GaAs channels. The differential magnetoresistance shows nonmonotonic behavior, a signature of electron pairing in the hydrodynamic regime. Theoretical models that account for both the influence of these interactions on shear stress relaxation and viscosity changes from electron heating show good agreement with the data. The nonlinear regime thus reveals how such correlated states govern the hydrodynamic behavior of the electron fluid. Our findings establish the nonlinear transport regime as a powerful probe for dissecting the complex interplay of correlated electron states and momentum relaxation in the hydrodynamic flow of an electron fluid.

\end{abstract}
\maketitle
\section{Introduction}
The discovery of the hydrodynamic regime in ultrapure conductors has transformed our understanding of electron transport, revealing a world where electrons flow collectively like a classical fluid~\cite{narozhny,hui}. This paradigm shift from single-particle kinetics to collective fluid dynamics is predicated on a simple hierarchy of scattering lengths: the electron-electron mean free path $l_{ee}$ must become the shortest scale in the system, notably smaller than the momentum-relaxing mean free path $l$ due to impurities and phonons. Under this condition, a wealth of striking phenomena has been experimentally confirmed, including the temperature-induced resistance reduction known as the Gurzhi effect \cite{gurzhi, dejong, gusev1, gusev2, levin}, giant negative magnetoresistance \cite{du, hatke, hatke2, shi2, alekseev1, raichev, gusev3, du2, gusev4}, the emergence of whirlpool-like current patterns manifesting as negative nonlocal resistance \cite{bandurin, torre, pellegrino2, levin2}, and Stokes-like flow around obstacles \cite{lucas, gusev5, gornyi, alekseev5, krebs}

Studies of electron hydrodynamics have so far concentrated primarily on the linear Stokes--Ohm regime, where transport phenomena remain relatively simple. A central objective, however, is to reach the nonlinear domain, where richer fluid-like effects can emerge. In classical fluids, nonlinearities typically become significant at high flow velocities, quantified by large Reynolds numbers. For electronic systems, this presents a major obstacle, since state-of-the-art devices generally operate at extremely low Reynolds numbers ($\text{Re} \sim 10^{-2}$), making nonlinear behavior difficult to access under standard conditions. This limitation has confined most investigations to the linear response regime, leaving the rich physics of nonlinear electron hydrodynamics largely unexplored.

Despite these challenges, recent theoretical advances suggest that strong magnetic fields and correlated electron interactions can provide alternative pathways to nonlinear behavior, even at modest flow velocities. The cyclotron motion of electrons in magnetic field can enhance memory effects and enable novel correlation mechanisms that dramatically alter the fluid's response. Recent theoretical work \cite{alekseev6} provides a compelling framework for such behavior, proposing that ``extended collisions'' of electron pairs on cyclotron orbits can introduce a non-local-in-time (memory) dependence into the shear stress~\cite{alekseev6}. This results in an effective non-Newtonian viscosity, which depends on the flow gradient, leading to predicted non-monotonic features in the nonlinear magnetoresistance.

This correlation-induced nonlinearity represents a distinctly quantum-mechanical pathway to non-Newtonian behavior that does not rely on conventional inertial effects. It means that the effective shear viscosity of the electron fluid becomes dependent on the velocity gradient, rather than remaining constant as in a Newtonian fluid. This leads to a nonlinear correction to the magnetoresistance, which we observe as an enhancement and reshaping of the differential magnetoresistance peak at low magnetic fields.

However, disentangling this mechanism from other nonlinearities, most prominently current-induced electron heating, remains a significant challenge. Heating alone can modify scattering times and mimic nonlinear flow effects, making the interpretation of experimental data complex. The competition and interplay between these different nonlinear mechanisms--correlation-driven memory effects versus heating-induced parameter renormalization--defines a new frontier in the study of electronic fluids.
\begin{figure}[ht]
\includegraphics[width=8cm]{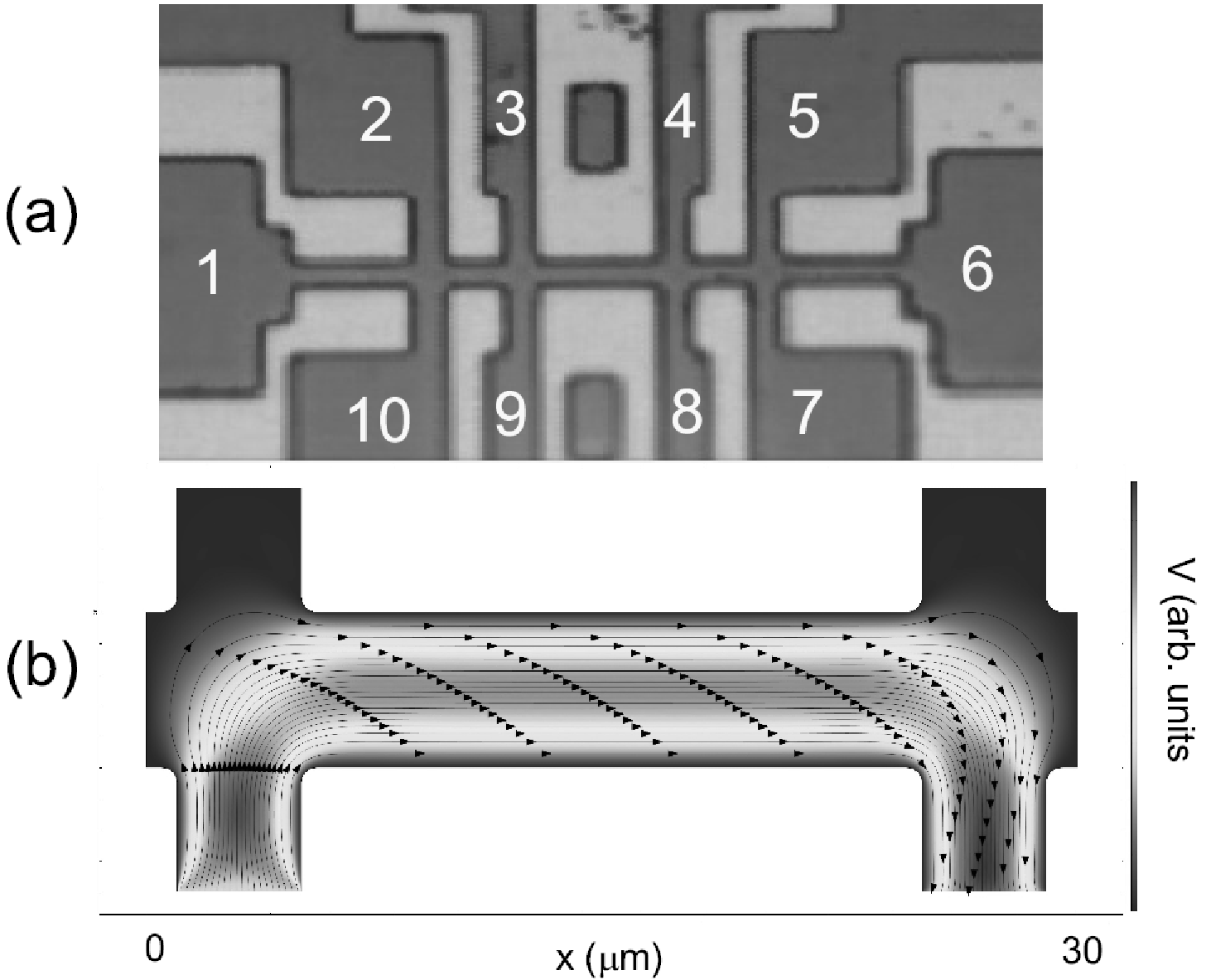}
\caption{\label{sample}(Color online)
Image of the central part of the Hall bar with 10 contacts. (b) Hydrodynamic velocity flow in configuration, when current is directed between contacts 9 and 8. Sketch of the velocity flow profile in a device within a channel of narrow width of $W = 6 \mu m$.}
\end{figure}
In this work, we overcome the low-Reynolds-number limitation and probe the heart of the nonlinear regime by studying magnetotransport in long, narrow channels of a high-purity GaAs two-dimensional electron gas. We observe a pronounced non-monotonic behavior in the differential magnetoresistance  at high currents, a clear signature of departure from standard hydrodynamics. By systematically comparing our data with theoretical models, we demonstrate that the observed nonlinearity is a composite effect. It requires a combination of the correlation-induced non-Newtonian viscosity, as described by model \cite{alekseev6}, and the more conventional effects of electron heating. Our findings not only confirm the existence of a non-Newtonian electron fluid but also establish a refined framework for interpreting nonlinear transport, where the interplay of memory effects and energy relaxation governs the ultimate hydrodynamic behavior.

Previous studies have examined the dependence of the differential resistance on the direct current in both macroscopic samples and high-mobility mesoscopic channels, revealing qualitatively similar behavior \cite{shi, studenikin}.
It should be noted that the pronounced negative magnetoresistance observed in macroscopic samples \cite{shi2} has been interpreted in some theoretical works as evidence of hydrodynamic electron flow in a lattice of oval defects \cite{alekseev1}, where such defects are assumed to act as effective momentum-relaxing scatterers. The existence and characteristics of oval defects in GaAs heterostructures were experimentally studied in Ref. \cite{bockhorn}, providing the material basis for this scenario, although the interpretation of transport data in terms of hydrodynamic flow through such defect lattices remains under discussion\cite{gornyi}.

Although a comparison between experimental data and theoretical models has been reported in Ref.~\cite{alekseev6}, the applicability of that model to the present system remains uncertain.
In particular, the possible influence of electron heating effects was not explicitly considered.
More recent work \cite{studenikin} have also not provided a detailed comparison with theoretical descriptions accounting for heating, nor with recently developed nonlinear transport theories.

A comprehensive analysis that incorporates these effects is therefore still lacking.
In the following, we perform such an analysis, taking into account both the electron heating and correlation-driven memory effects, and compare the results with our experimental data obtained from narrow channels.
\section{Experimental results}

\begin{figure}[ht]
\includegraphics[width=8cm]{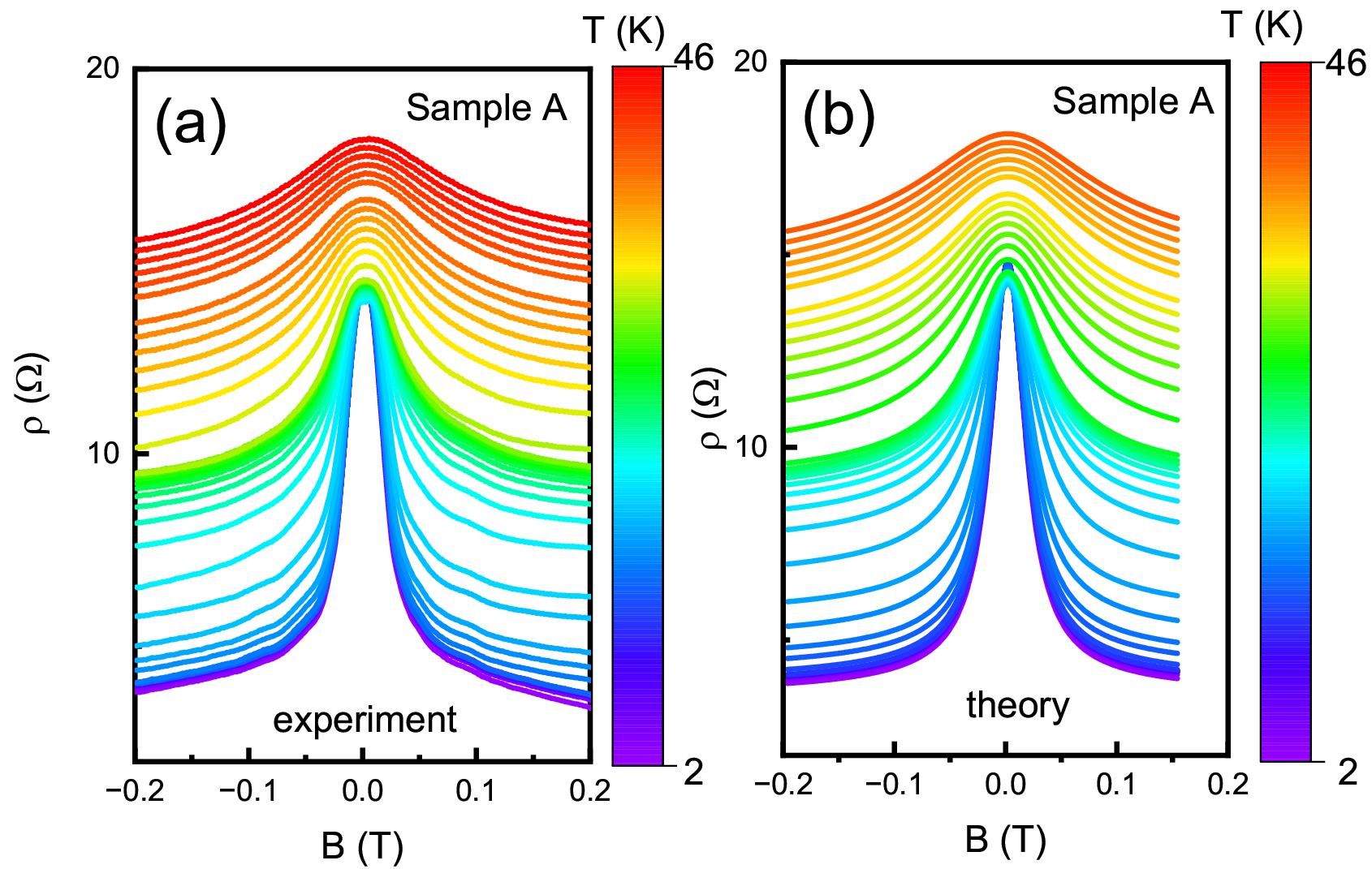}
\caption{\label{Tdependence}(Color online)
(a) Temperature-dependent magnetoresistivity of narrow GaAs channel, sample A. (b) Temperature-dependent magnetoresistivity calculated from equation  ~\eqref{magnetoresistance}, sample A .}
\end{figure}
The devices investigated in this study were fabricated from high-quality GaAs quantum wells with a nominal width of \SI{14}{nm} and an electron density of approximately
\(\ensuremath{9.0\times10^{11}\,\mathrm{cm^{-2}}}\) at \SI{4.2}{K}.
The macroscopic sample exhibited an electron mobility of
\(\ensuremath{2\times10^{6}\,\mathrm{cm^{2}/Vs}}\).
For transport measurements, we employed a Hall-bar geometry optimized for multiterminal experiments.
The structure consisted of three consecutive segments with lengths of \SI{6}{\micro\meter}, \SI{20}{\micro\meter}, and \SI{6}{\micro\meter},
each having a width of \SI{6}{\micro\meter}, and incorporated ten voltage probes.
Ohmic contacts to the two-dimensional electron gas (2DEG) were formed by annealing indium (In) layers deposited on the GaAs surface.
An optical micrograph of the device with numbered contacts is presented in  Fig.~\ref{sample}a.
Two nominally identical samples were studied under identical measurement conditions.

Measurements were conducted in a variable temperature insert (VTI) cryostat using a standard lock-in detection technique to record the longitudinal resistance.
To minimize electron overheating, a low-frequency alternating current (AC) in the range of \SIrange{0.1}{1}{\micro\ampere} was applied.
Simultaneously, a direct current \(I_{\mathrm{dc}}\) was superimposed through the same leads to determine the differential resistance, defined as $r_d = r_{xx} = \frac{dV_{xx}}{dI_{\mathrm{dc}}}$.
The experiments were carried out using an H-patterned sample configuration designed to enhance hydrodynamic transport effects~\cite{gusev1}.
In this configuration, the current \(I\) was driven between contacts 9 and 8, while the voltage \(V\) was measured across probes 3 and 4, yielding $R = R^{3,4}_{9,8} = \frac{V_{3,4}}{I_{9,8}}$
(see Fig.~\ref{sample}b).
The primary focus of this work is the investigation of magnetoresistivity and zero-field resistivity as functions of temperature and direct current \(I_{\mathrm{dc}}\).

Fig.~\ref{Tdependence}a presents the magnetic-field dependence of the resistivity, $\rho = \frac{W}{L} \times R$, at various temperatures for the sample A. A pronounced negative magnetoresistivity, characterized by \(\rho(B) - \rho(0) < 0\), is observed and well described by a Lorentzian profile.
With increasing temperature, both the amplitude of the negative magnetoresistivity and its sharpness decrease, while the zero-field resistivity increases.
Comparable magnetoresistance behavior was observed in the second sample B (not shown), confirming the reproducibility and robustness of the transport phenomena reported here.

\begin{figure}[ht]
\includegraphics[width=8cm]{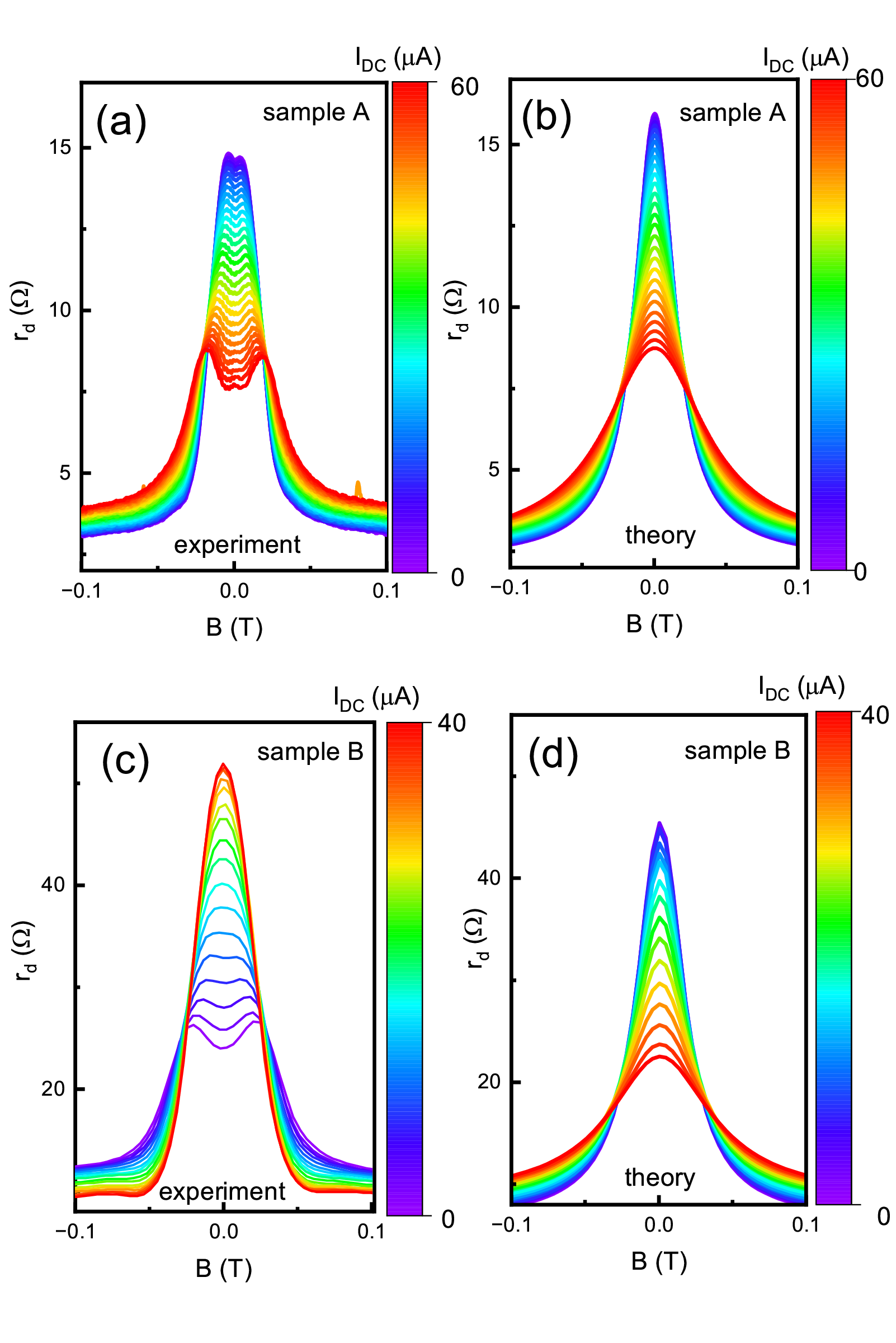}
\caption{\label{MRbias}(Color online) Evolution of differential magnetoresistance $r_d$ with increasing dc bias for sample A (a) and sample B (c). Magnetoresistivity calculated from equation  ~\eqref{magnetoresistance} for sample A (b) and sample B (d). }
\end{figure}
\begin{figure}[ht]
\includegraphics[width=8cm]{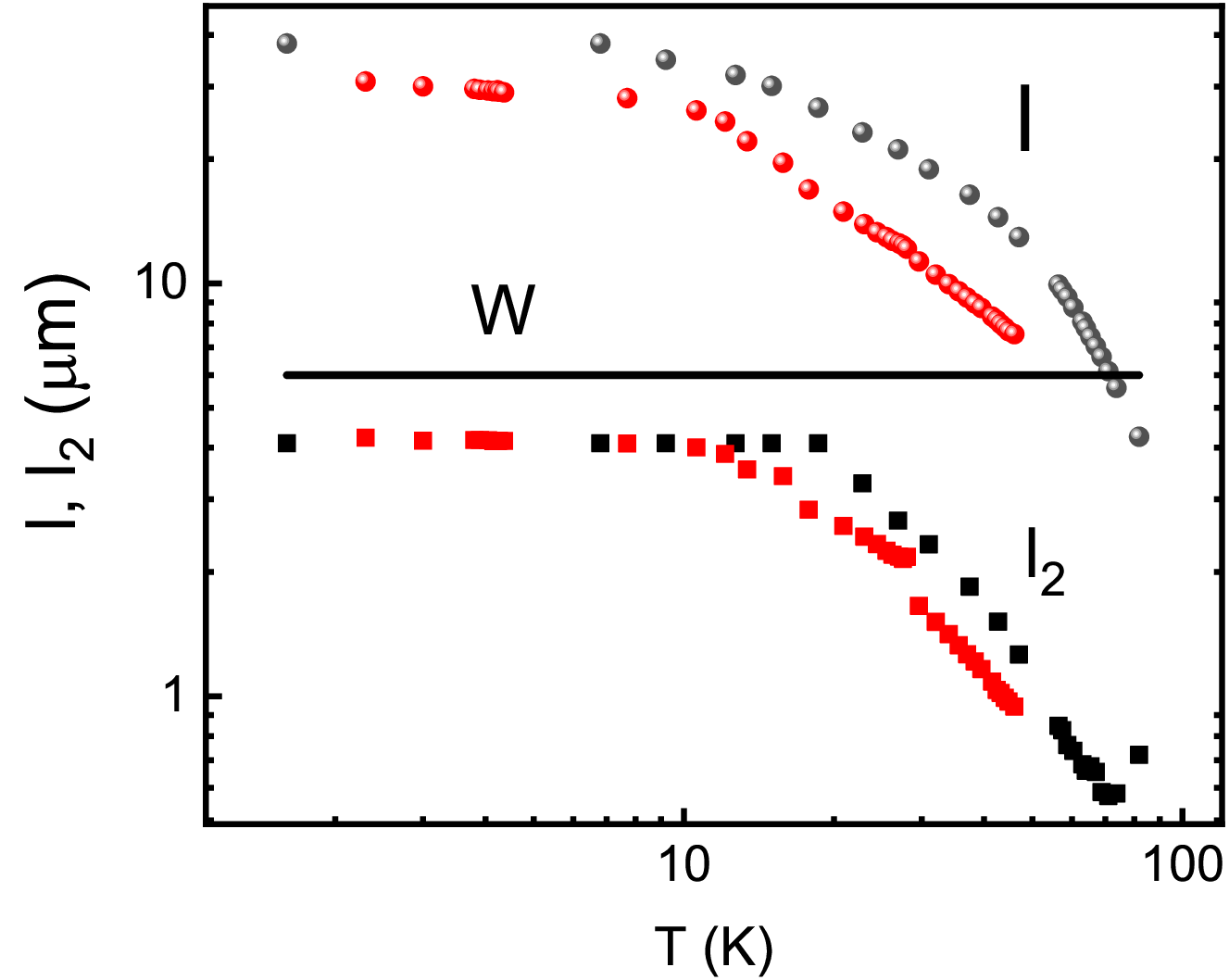}
\caption{\label{length}(Color online) The characteristic lengths $l$, $l_2$  as a function of temperature for two reference samples; (black -Sample A, red- sample B). Horizontal line- the width of the sample W.}
\end{figure}
\begin{figure}[ht]
\includegraphics[width=8cm]{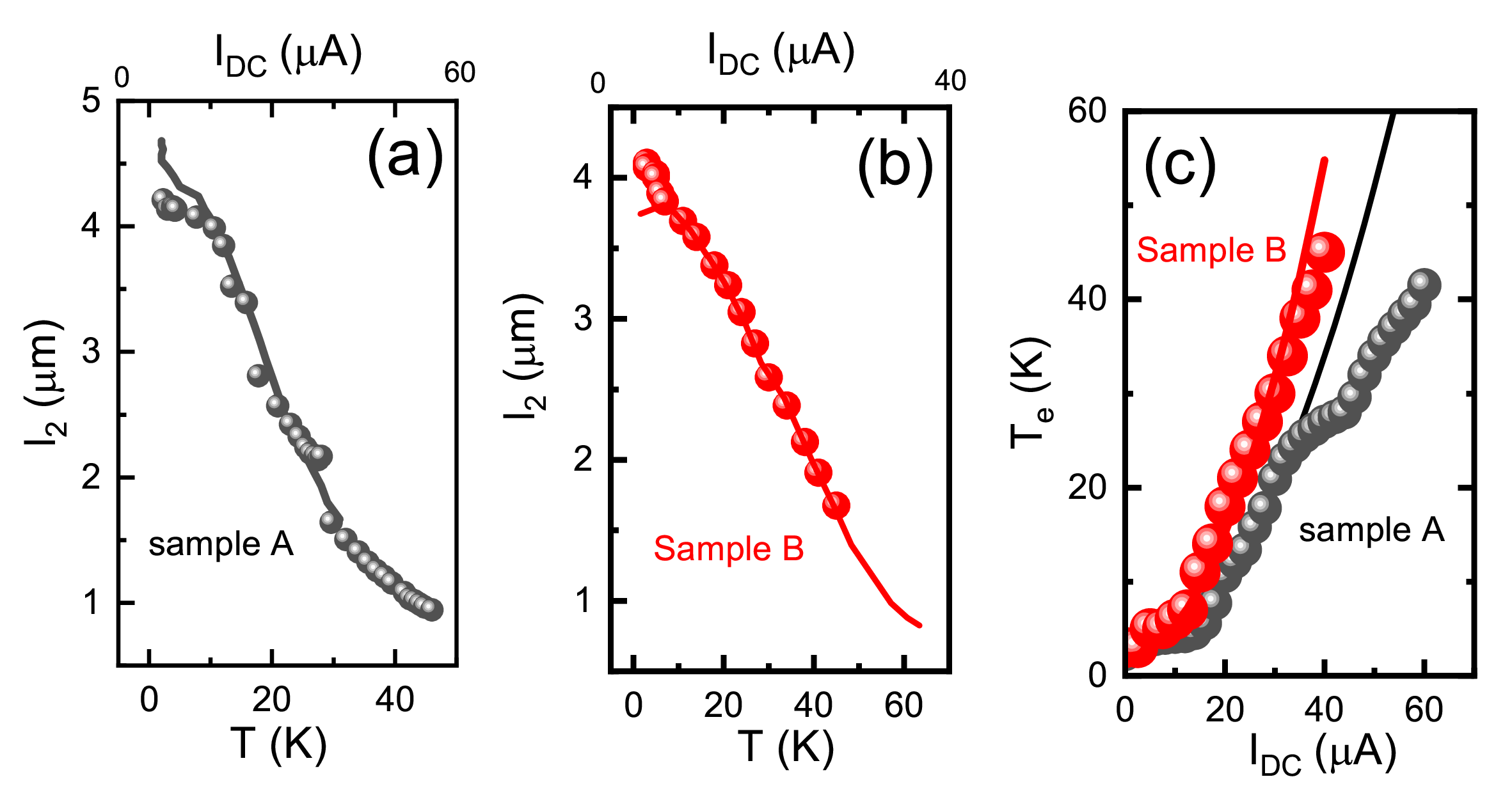}
\caption{\label{bias}(Color online) (a) Comparative analysis of the fitting parameter $\l_{2}$ for Sample A, showing its dependence on temperature (circles) and DC bias current (solid line).
(b) Comparative analysis of the fitting parameter $\l_{2}$ for Sample B, showing its dependence on temperature  (circles) and DC bias current ( solid line).
(c) The extracted electron temperature as a function of DC current for Samples A and B, as determined from the analyses presented in (a) and (b).The solid lines show the expected trend for a quadratic dependence ($\sim I_{DC}^2$). }
\end{figure}

In addition, we measured the differential resistance \(r\) as a function of the direct current \(I_{\mathrm{dc}}\) at a fixed low temperature of \(T = \SI{2.5}{K}\).
The set of experimental curves obtained for both samples~A and~B is presented in Fig.~\ref{MRbias}(a,c).
In contrast to the temperature-dependent measurements, the magnetoresistance exhibits a nonmonotonic dependence on the magnetic field.
At certain values of \(B\), the resistance displays a distinct peak whose position shifts toward higher magnetic fields with increasing \(I_{\mathrm{dc}}\).
Simultaneously, the resistivity at zero magnetic field decreases with increasing \(I_{\mathrm{dc}}\), and the Lorentzian-shaped magnetoresistivity broadens correspondingly.
These observations indicate that the application of a finite direct current significantly modifies the electron transport regime, highlighting the nonlinear character of the magnetoresistive response.

\section{Comparison with Theory: Influence of Heating Effects}
To provide a qualitative comparison with experimental results obtained from obstacle-free samples, we utilize a model established in prior studies, originally developed to characterize Poiseuille flow in the presence of a magnetic field \cite{alekseev1}.

This model expresses the resistivity as arising from two primary components. The first originates from ballistic phenomena or scattering at boundaries and defects, and the second from viscous effects \cite{alekseev1}.
\begin{equation}
\rho(B)= \frac{m}{e^{2}n}\left(\frac{1}{\tau}+\frac{1}{\tau^{*}}\right)
\end{equation}
Here, $1/\tau$ is the scattering rate from static disorder, while $m=0.63m_{0}$ and $n$ are the effective mass and carrier density, respectively, with $m_0$ being the electron rest mass. The viscous relaxation time is given by $\tau^{*} = W^{*2}/(12\eta)$, where the viscosity is $\eta = v_{F}^{2}\tau_{2}/4$. The parameter $W^{*}$ denotes the effective sample width, which equals the physical width $W$ for zero-slip boundary conditions. The relaxation time $\tau_{2}$ describes the decay of shear stress due to electron-electron collisions. The subscript "2" indicates that this viscosity component is associated with the relaxation of the second harmonic of the electron distribution function \cite{alekseev1}.

For a comprehensive description under a magnetic field, the theory employs a field-dependent viscosity tensor to derive the resistivity tensor:
\begin{equation}
\label{magnetoresistance}
\rho(B)=\left(\frac{m}{e^{2}n\tau}\right)\frac{1}{1-\tanh(\xi)/\xi}
\end{equation}
The dimensionless Gurzhi parameter here is $\xi = \xi_{0} \sqrt{1 + (2l_{2}/r_c)^2}$, with $\xi_{0} = W/l_{G}$. The Gurzhi length is $l_{G} = \sqrt{l_{2}l}$, where $l_{2} = v_{F} \tau_{2}$, $l = v_{F} \tau$, and the cyclotron radius is $r_{c} = v_{F}/\omega_{c}$. The cyclotron frequency is $\omega_{c} = eB/mc$. The relaxation rate for shear viscosity combines contributions from electron-electron interactions and impurities:
\begin{equation}
1/\tau_{2}(T) = 1/\tau_{2,ee}(T) + 1/\tau_{2,imp}
\end{equation}
Similarly, the momentum relaxation rate includes phonon and impurity scattering:
\begin{equation}
\label{tau}
1/\tau(T) = 1/\tau_{0,ph}(T) + 1/\tau_{0,imp}
\end{equation}
Here, $1/\tau_{0,ph} = B_{ph}T$ accounts for phonon scattering, while $1/\tau_{0,imp}$ represents scattering from static disorder, which is different from the relaxation time for the second harmonic \cite{alekseev1}.

The linear-in-temperature form of the phonon scattering rate is well established for acoustic-phonon scattering in a degenerate two-dimensional electron gas.
The coefficient $B_{\mathrm{ph}}$ is material- and sample-dependent and incorporates the microscopic electron--phonon coupling mechanisms, including deformation-potential  and piezoelectric  interactions, as well as sound velocities, mass density, and screening effects. Its value can be estimated from microscopic theory or extracted experimentally from temperature-dependent transport measurements.
Experimentally, the linear temperature dependence of the mobility (and hence the scattering rate) in high-mobility GaAs 2DEGs was demonstrated in Ref.~\cite{harris}. A detailed theoretical analysis was provided in Ref.~\cite{sarma}. In the context of viscous electron hydrodynamics, the model of Ref.~\cite{alekseev1} employed the same linear form for $1/\tau_{0,\mathrm{ph}}$ with $B_{\mathrm{ph}} \approx 10^{9}\,\mathrm{s^{-1}K^{-1}}$, which is consistent with the values extracted in the present work.
In our mesoscopic samples, transport reflects a combination of hydrodynamic and ballistic contributions [Eqs.~(1) and (2)], which leads to some variation in the effective $B_{\mathrm{ph}}$ between samples A and B. This variation likely reflects the sensitivity of the phonon scattering rate to confinement, boundary conditions, and the spatial structure of the electron flow in narrow channels.

We subsequently fit both the magnetoresistance data and the zero-field resistivity $\rho(T)$ using three adjustable parameters: $\tau(T)$, $\tau_{2}(T)$, and the effective width $W^{*}$ (Fig.~\ref{MRbias}(b,d)).
 Fig.~\ref{length} represents temperature dependence of the characteristic lengths $l$ and $l_2$  extracted from experiments on the two reference samples.  The analysis of the magnetoresistance data yields information on electron-electron interactions and disorder-induced relaxation. For direct comparison with theory, the parameters \( 1/\tau_{2,imp} \), \( 1/\tau_{0,imp} \), \( A_{ee} \), and \( B_{ph} \) are extracted, as summarized in Table 1. Using Eq. (3), the electron-electron relaxation rate is given by:
\begin{equation}
\frac{\hbar}{\tau_{2,ee}} = A_{ee} \frac{(kT)^2}{E_F}
\end{equation}

The results show that all relaxation rates follow universal scaling laws: \( 1/\tau_{2,ee} \propto T^2 \) and \( 1/\tau \propto T \).

To facilitate comparison with theoretical predictions, we used the parameters \( \frac{1}{\tau_{2,imp}} \), \( \frac{1}{\tau_{0,imp}} \), \( A_{ee} \), and \( B_{ph} \), as listed in Table 1.
\begin{table}[ht]
\caption{\label{tab1} Fitting parameters of the electron system. Parameters are defined in the text.}
\begin{ruledtabular}
\begin{tabular}{lcccccc}
& Sample & $1/\tau_{2,imp}$&$1/\tau_{0,imp}$ & $A_{ee}$ & $B_{ph}$ &$W^{*}$   \\
& &  $(10^{11} 1/s)$ & $(10^{10} 1/s)$ &   & ($10^{9} 1/sK$) & $\mu m$  \\
\hline
& A &  $1.0$  & $1.0$ & $0.5$ &  $0.9$ & 11.3\\
& B &  $1.0$  & $0.8$ & $0.6$ &  $0.5$ & 6\\
\end{tabular}
\end{ruledtabular}
\end{table}

Based on these results, we can evaluate the conditions for observing hydrodynamic electron flow in our samples. The hydrodynamic regime generally requires that the electron-electron scattering mean free path, \(l_{ee}\), is the shortest length scale, specifically shorter than the sample width \(W\), which itself must be smaller than the momentum-relaxing mean free path \(l\). This establishes the hierarchy \(l_{ee} < W < l\), which can be seen in the Figure Fig.~\ref{length}. A more precise criterion involves the relaxation of the second harmonic of the electron distribution, characterized by the rate \(1/\tau_2 = v_F / l_2\), which must exceed the electron-electron scattering rate \cite{alekseev1}. This relaxation process can be influenced by impurities, resulting in a finite shear stress relaxation rate even at very low temperatures due to disorder \cite{alekseev7}.

The table further reveals that in sample A, the effective width exceeds the geometric width. This is consistent with theoretical models that account for a finite slip length at the boundaries \cite{gromov}. Using the relation  \( W^{*2} = W(W + 6l_s) \), we estimate a slip length $l_s$ of approximately 4–5 $\mu m$. A finite slip length alters the flow profile from a classic parabolic shape to a "cut parabola," effectively extending the channel's hydrodynamic width \cite{gusev2, kiselev, raichev2}. While the slip length is a complex parameter highly dependent on boundary conditions and warrants independent study, such an investigation falls outside the scope of this work.

For our samples, the relationship \(l_2 < W < l\) ( Fig.~\ref{length}), or in terms of rates, \(1/\tau_2 > v_F/W > 1/\tau\), is satisfied. This confirms that the system is in the hydrodynamic regime down to the base temperature of \(T = 4.2\,\text{K}\). Additionally, in mesoscopic systems, the momentum scattering rate \(1/\tau\) is typically suppressed compared to bulk materials because of the prevalence of boundary scattering and geometric confinement effects.

 A more stringent requirement, \(1/\tau_{2,ee} > v_F/W > 1/\tau\), which specifically uses the electron-electron contribution to the second harmonic relaxation, is fulfilled within an intermediate temperature window of \(20\,\text{K} < T < 60\,\text{K}\).

Furthermore, we compare the experimental results for nonlinear transport under the assumption that the broadening of the magnetoresistivity peak and its reduction in amplitude originate from electron heating effects, which significantly influence electron–electron (e–e) scattering.
It is important to note, however, a distinct difference in the behavior of the resistivity at zero magnetic field as functions of temperature and direct current:
while \(\rho(B = 0)\) increases with rising temperature \(T\), the differential resistance \(r(B = 0)\) decreases with increasing \(I_{\mathrm{dc}}\).
We interpret this difference as evidence that the applied direct current primarily increases the electron temperature, without substantially affecting the lattice temperature.
As a result, electron–phonon (e–ph) scattering does not contribute in this case, and only the second term in equation \eqref{tau}  should be considered.
In contrast, when the bath temperature increases, both terms in Eq.\eqref{tau}  become relevant, since the e–ph scattering time decreases.
This distinction has also been discussed in previous studies of hydrodynamic effects in GaAs channels \cite{dejong}, where the Gurzhi effect was not directly observed but inferred indirectly through the application of a finite direct current.

The results of the comparison with equation \eqref{magnetoresistance}  are presented in Fig.~\ref{MRbias} for sample A.
It can be seen that while the Lorentzian peak shape and amplitude are satisfactorily described by the model~\cite{alekseev1}, the features near zero magnetic field are not captured by it.
From the fitting procedure based on this model, we also extract the characteristic parameters \(l_2\) and \(l\).
Since only \(l_2\) exhibits temperature dependence, we use this relation for electron thermometry.
Figs.~\ref{bias}(a,b) display both dependencies, \(l_2(T)\) and \(l_2(I_{\mathrm{dc}})\), with the temperature and current scales adjusted accordingly.
By comparing these two dependencies, we determine the electron temperature as a function of direct current, \(T_{\mathrm{e}}(I_{\mathrm{dc}})\), shown in Fig.~\ref{bias}c for both samples.
At low direct currents, \(T_{\mathrm{e}}\) follows the expected quadratic dependence on \(I_{\mathrm{dc}}^2\). At higher dc currents, deviations from the quadratic dependence for sample A become apparent. This behavior reflects the strongly temperature-dependent nature of the electron–phonon energy-loss rate in GaAs two-dimensional electron gases. As the electron temperature increases, energy relaxation via acoustic phonon emission becomes increasingly efficient, leading to a sublinear, saturation-like dependence of $T_{e}$ on the dissipated power, consistent with previous experimental observations \cite{proskuryakov}. The fact that such deviations are visible in sample A but not in sample B is naturally attributed to sample-specific differences (e.g., geometry and boundary conditions), which affect the onset of this crossover and the resulting current–temperature relation.
We find that the electron temperature can reach values up to approximately \SI{40}{K}, confirming that under high-current conditions the system enters the hydrodynamic transport regime, where comparison with hydrodynamic models becomes physically relevant.

\section{Comparison with Theory in the Nonlinear Magnetotransport Regime}
\begin{figure}[ht]
\includegraphics[width=8cm]{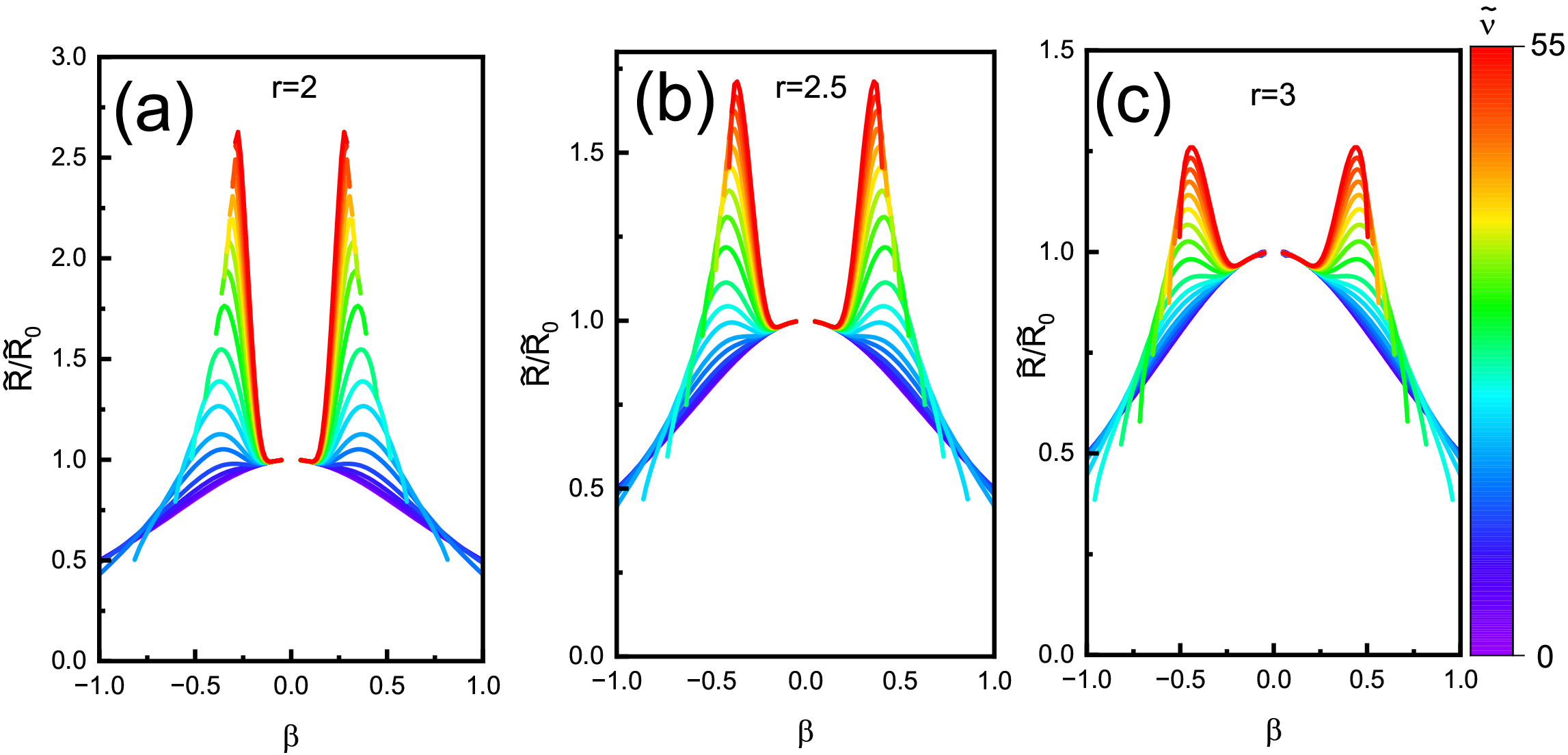}
\caption{\label{biastheory}(Color online) Evolution of differential magnetoresistance $\widetilde{R}$ with increasing dc bias calculated from equation  ~\eqref{resistance}  for different parameters $r$ :2 (a), 2.5 (b), 3 (c). Dimensionless bias parameter $\widetilde{\nu}$ is defined in the text.  }
\end{figure}
\begin{figure}[ht]
\includegraphics[width=8cm]{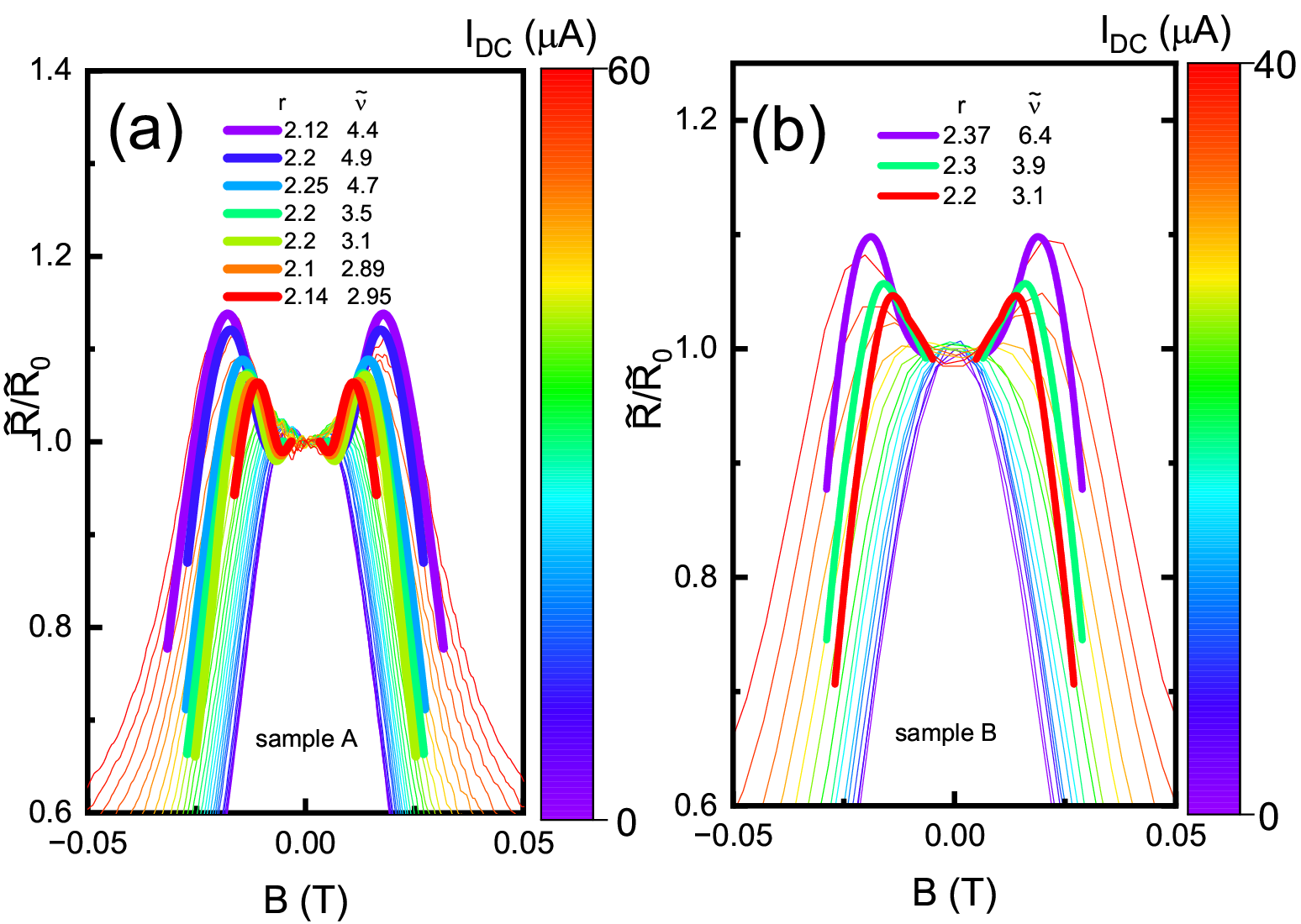}
\caption{\label{biasresistance}(Color online) Evolution of differential relative  magnetoresistance $\widetilde{R}/\widetilde{R_{0}}$ with increasing dc bias at low magnetic field. Thick solid lines represent theoretical curves calculated from \eqref{resistance}  for different values of $r$ and $\widetilde{\nu}$. as indicated in the legend.}
\end{figure}
\begin{figure}[ht]
\includegraphics[width=8cm]{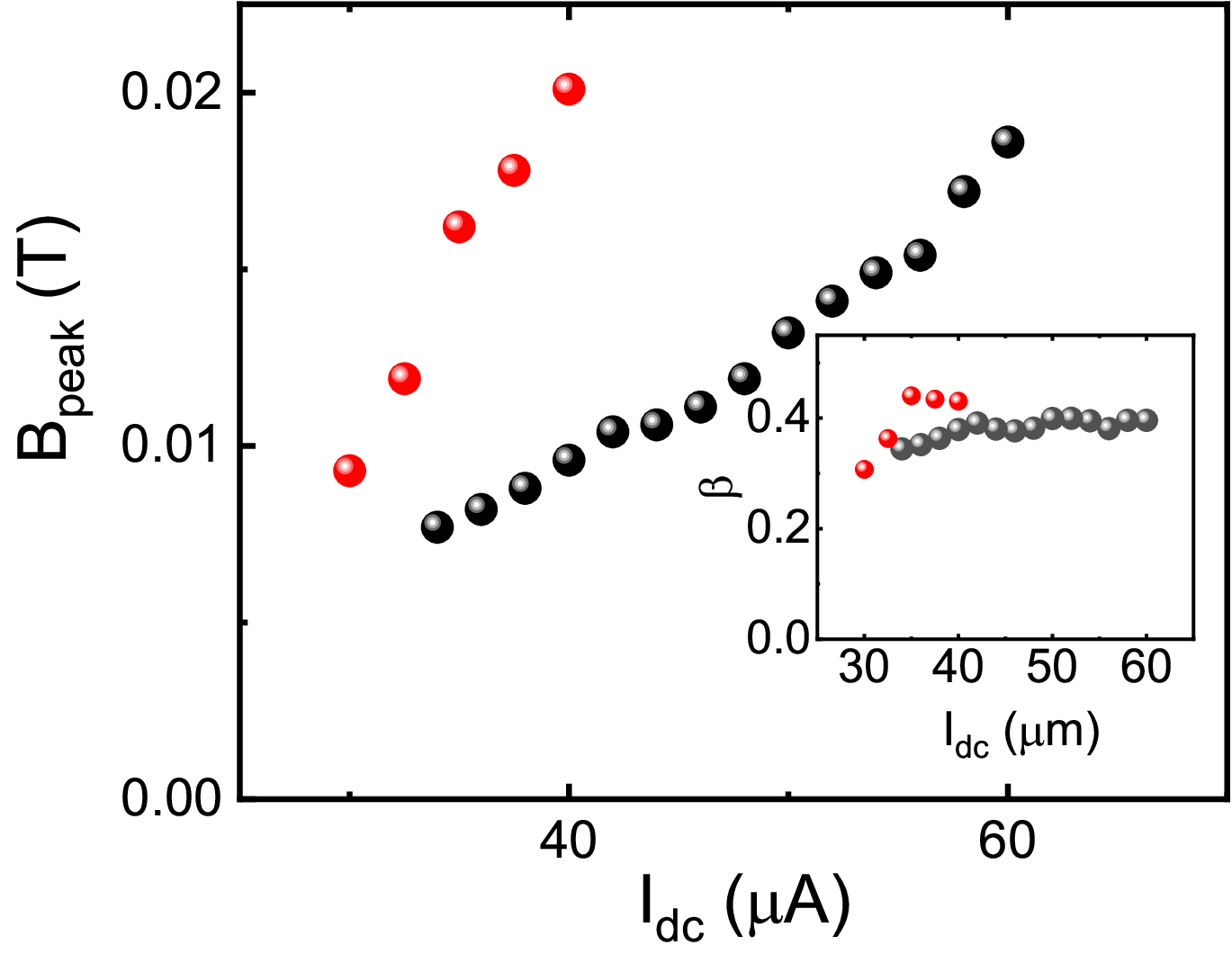}
\caption{\label{peak}(Color online) Position of the magnetoresistivity peak as a function of the applied direct current \(I_{\mathrm{dc}}\). The solid line represents a parabolic fit.
\textit{Inset:} the dimensionless parameter \(\beta = 2\omega_{c}\tau_{2,ee}\) calculated at the peak position, together with the corresponding relaxation time determined for each heating current.
Black points-sample A, red points-sample B.}

\end{figure}
The theoretical model \cite{alekseev6} demonstrated that the nonlinear
magnetohydrodynamic transport of an electron fluid arises from
pair correlations formed during, so called, "extended collisions".
These collisions occur as electrons traverse their cyclotron orbits,
introducing a nonlinear memory effect into the relaxation of shear stress.
In the regime of slow flows, this memory effect manifests as a viscosity
that depends on the velocity gradient, thereby causing the electron fluid
to deviate from Newtonian behavior.

This mechanism establishes a direct link between microscopic electron
dynamics and macroscopic hydrodynamic properties: the magnetic field not only
governs the individual cyclotron motion of electrons but also modulates the
overall transport coefficients of the fluid through these orbital correlations.

The key dimensionless parameter governing this nonlinearity is defined as
\[
    r = \frac{4 \pi \tau_2^{*}}{\tau_q}.
\]
Here, $\tau_2^{*}$ denotes the lifetime of the paired correlation state, a collective
effect in which electrons move in a correlated fashion, reinforcing hydrodynamic
behavior \cite{alekseev6}. Notes that In contrast, $\tau_q$ represents the quasiparticle scattering time,
which characterizes conventional momentum relaxation arising from uncorrelated
electron--electron or electron--impurity interactions. The ratio $r$ thus expresses
the balance between collective hydrodynamic enhancement and individual scattering
processes, encapsulating the degree to which the electron system exhibits
nonlinear, non-Newtonian transport behavior.

The dimensionless differential resistance $\widetilde{R}$ in the model ~\cite{alekseev6}
was evaluated numerically as a function of two key parameters: the ratio $r$, which quantifies the relative significance
of pairwise electron correlations compared to other scattering processes, and the scaled nonlinearity parameter
$\widetilde{\nu} = C_{\nu} \nu$, which depends on the applied electric field, $C_{\nu} = 2^{10}\pi^{2}C_{\alpha}$. It is expected that $C_{\alpha} \sim 1$.
The intrinsic dimensionless  nonlinearity parameter is given by
\begin{equation}
\label{nonlinear}
\nu = \left( \frac{l_2}{a_B} \frac{e E W}{m v_F^2} \right)^2,
\end{equation}
where  the effective Bohr radius is defined as $a_B = \frac{4 \pi \epsilon \epsilon_0 \hbar^2}{m_e e^2}$.
Subsequently, the resistance $\widetilde{R}$ was computed as a function of the magnetic field, expressed  through the additional dimensionless parameter $\beta = 2 \omega_c \tau_2$.
Additional parameters are defined:
\begin{equation}
\label{nu}
\nu'(\beta) = f(\beta) \, \widetilde{\nu}, \quad \text{with } f(\beta) = \frac{e^{-r/\beta}}{\beta^4}
\end{equation}
The dimensionless differential resistance $\widetilde{R}$ is given by equation \cite{alekseev6} :
\begin{equation}
    \label{resistance}
    \widetilde{R}(\beta, \widetilde{\nu}) = \frac{1}{\widetilde{I} + 2\widetilde{\nu} \, \frac{d\widetilde{I}}{d\widetilde{\nu}}}
    \end{equation}
where dimensionless current is found from the integral
$\tilde{I} = \tilde{I}(\nu, \beta) = \int_{W/2}^{-W/2} d\widetilde{y} \, \tilde{V}(\widetilde{y})$. Variable  $\widetilde{y}(u)$ and  function $\tilde{V}(\widetilde{y})$ are found from other dimensionless parameters:
\begin{align}
\widetilde{y}(u) &= -\frac{(1-\nu'u^2)u}{\beta^2+(1-\nu'u^2)^2}, \quad
\widetilde{V}(u) = \widetilde{V}_1 - \frac{Y(u)}{\nu'}, \\
Y(u) &= \frac{\beta^2+1-\nu'u^2}{\beta^2+(1-\nu'u^2)^2} + \frac{\ln[\beta^2+(1-\nu'u^2)^2]}{4}.
\end{align}
The parameter $u$ varies within $(-u_0, u_0)$, where $u = \pm u_0$  corresponds to the sample edges at $\widetilde{y} = \pm 1/2$.  The constant $\widetilde{V}_1$ is determined from the boundary conditions $\widetilde{V}(\pm u_0) = 0$.

Equation \eqref{resistance} is compared with our experimental results, leading to several observations. Pair correlations produce a non-monotonic dependence of $\widetilde{R}$ on the magnetic field, manifesting as a peak or minimum in the resistance-field curve, as clearly seen in Fig.~\ref{biastheory}. The observed peak arises from the competition between the Lorentz force (induced by the magnetic field) and viscous effects, which are amplified by pair correlations.
We calculated the relative dimensionless differential resistance
\(\widetilde{R}(\beta, \widetilde{\nu}) / \widetilde{R}(\beta, \widetilde{\nu} = 0)\)
for various values of the parameters \(r\) and \(\widetilde{\nu}\), and the resulting dependencies are presented in Fig.~\ref{biastheory}.
It can be seen that the differential resistance exhibits a pronounced peak at a certain magnetic field.
The position of this peak depends on the parameter \(r\), but remains essentially independent of \(\widetilde{\nu}\), i.e., the applied direct current.
In contrast, the relative peak amplitude increases significantly with decreasing \(r\) for a given \(\widetilde{\nu}\), indicating that the parameter \(r\) strongly influences the nonlinear response of the system.

The theoretical description predicts a strong sensitivity of the nonlinear magnetoresistance to the dimensionless parameter (r), which characterizes correlation-induced memory effects in electron–electron scattering. In our experiment, the data are consistently described by values of ($r \approx 2.1\text{–}2.4$). Calculations for larger (r) indicate that the non-monotonic magnetoresistance peak is progressively suppressed and eventually disappears, signaling a crossover toward a more Newtonian regime with effectively constant viscosity. This trend is illustrated both by the theoretical results of Ref. \cite{alekseev6} and by our simulations in Fig.~\ref{biastheory}, where increasing (r) leads to a reduction of the peak amplitude. Very large values of (r) correspond to a regime in which momentum-conserving collisions become weakly correlated, which may be difficult to realize in conventional high-mobility two-dimensional electron systems. Exploring the behavior over a broader range of (r) would require access to regimes with enhanced interaction effects or strong velocity-dependent correlations and is left for future experimental and theoretical studies.

Note that in Refs.~\cite{alekseev6, alekseev7}, the authors postulate that the stress relaxation time $\tau_2$—which governs viscosity and the Poiseuille profile—and the pair correlation time $\tau_2^{*}$—which determines the strength of memory effects and non-Newtonian behavior—are related but distinct concepts. For simplicity, we assume $\tau_2 \sim \tau_{2}^{*}$.

Taking this into account, we fitted the nonlinear behavior of the magnetoresistance in the vicinity of zero magnetic field. Figure~\ref{biasresistance} presents a comparison between the experimentally measured nonlinear curves and the differential relative magnetoresistance, \(\widetilde{R}/\widetilde{R_{0}}\), calculated from Eq.~\eqref{resistance} for both samples. A good agreement is observed between the experimental data and the theoretical predictions in terms of the peak position and its relative height. However, the theoretical curve exhibits a more abrupt decay on the high-field side of the peak compared to the experimental data. We attribute this discrepancy to the instability of the differential-equation solution at high magnetic fields under large dc currents. This behavior is also evident in Fig.~\ref{biastheory}, where the curves corresponding to higher dc currents are confined to lower magnetic fields compared to those at smaller dc currents.

Therefore, one can see that non-Newtonian behavior refers to the breakdown of a constant (Newtonian) viscosity due to velocity-dependent electron–electron correlations in the nonlinear hydrodynamic regime. This effect manifests itself experimentally as a bias-dependent modification of the hydrodynamic contribution to the magnetoresistance. In particular, at the highest dc currents the amplitude of the non-monotonic magnetoresistance peak increases by up to approximately $25\%$ relative to the low-bias, Newtonian response, as shown in Fig.~\ref{biasresistance}. This enhancement provides a direct quantitative measure of the non-Newtonian character of electron flow in our experiment.

 In our comparison, we used two fitting parameters, \(r\) and \(\widetilde{\nu}\). However, to explain the shift of the peak with increasing dc current, we considered the heating effect, which leads to an increase in the parameter \(\beta\ \sim\tau_{2,ee}\) calculated at the peak position. To examine this effect in more detail, we extracted the peak position as a function of the dc current and plotted the results in Fig.~\ref{peak}. An increase in DC current results in a shift of the position toward higher magnetic fields.

Furthermore, we calculated the parameter \(\beta\) for these magnetic fields, taking into account the reduction of the relaxation time \(\tau_{2,ee}\) with increasing current, as determined from the heating experiments shown in Figs.~\ref{bias}(a,b). The inset of Fig.~\ref{peak} demonstrates that the parameter \(\beta\) remains independent of the dc current, indicating that the parameter \(r\) also remains constant at different current levels, as predicted by the theory. This confirms our assumption that the peak shift observed in Fig.~\ref{peak} is entirely due to heating effects, whereas the increase in peak amplitude originates from nonlinear hydrodynamic effects.

Furthermore, we estimated the electric field and analyzed the nonlinear experimental conditions. Remarkably, the fitting parameter \(\widetilde{\nu} = 4\text{--}6\), obtained from the comparison with the experimental curves (Figs.~\ref{biasresistance}(a,b)), is found to be much smaller than the value calculated from the applied dc current, which reaches \(\widetilde{\nu} \approx 10^{4}\). This discrepancy indicates that the experimentally observed nonlinearity occurs at significantly lower effective values of \(\widetilde{\nu}\) than expected from theoretical estimates. It is worth noting that a similar inconsistency has been reported previously in Ref.~\cite{alekseev7}.

We can assume here that the small $C_\alpha$ likely stems from an overestimation in the qualitative estimate, where the characteristic distance $\lvert r_{1} - r_{2} \rvert$ between paired electrons is assumed to be the cyclotron radius $R_c$. This leads to $\alpha = C_{\alpha} P(\omega_{c}) \left( \frac{R_{c}}{a_{B}} \right)^{2} \frac{1}{\tau_{2}}$, therefore the temperature dependent correction to the relaxation time is
$T_{\Delta}= C_{\alpha} P(\omega_{c}) \left( \frac{R_{c}}{a_{B}} \right)^{2}
T_c \left( \frac{2\pi}{\omega_{c} \tau_{2}} \right)$, resulting to $f(\beta) = P(B)/\beta^{4}$  and large value of $C_{\nu}$ \cite{alekseev6}, $T_c$ is cyclotron period. Note that both the quantities \(1/\tau_{2}\) and \(\alpha\) are proportional to the probability
\(P(B) = e^{-T_c / \tau_{q}}\), which represents the likelihood that a particle in a pair
completes one cyclotron orbit without colliding with other (“third”) particles. However, for screened Coulomb interactions in GaAs, which are short-range (\(\sim a_{B}\)), only nearby pairs  \(\lvert r_{1} - r_{2} \rvert \sim a_{B}\) contribute significantly to changes in the scattering cross-section,  since more distant pairs interact only weakly. Therefore, the corrected expression for the coefficient becomes
$\alpha^{*} = C_{\alpha} P(\omega_{c}) \frac{1}{\tau_{2}}$. Accordingly, the corresponding temperature-dependent term is $T_{\Delta}^{*} = C_{\alpha} P(\omega_{c})
T \left( \frac{2\pi}{\omega_{c} \tau_{2}} \right)$, therefore $f^{*}(\beta) = \frac{P(B)}{\beta^{2}} = \frac{e^{-r / \beta}}{\beta^{2}}$. The revised nonlinear parameter becomes $C_\nu=(2\pi)^2 C_{\alpha}\approx 39.5C_{\alpha}$. Comparison with experiments allows us to estimate the coefficient  \(C_{\alpha} \sim 0.1\text{--}0.2\), which brings the experimental results  into much closer agreement with the model, although further numerical
analysis is still required.

As a final remark, we note that in this paper we focus on longitudinal magnetoresistance. The Hall response in the hydrodynamic regime is governed by mechanisms distinct from those controlling longitudinal transport and is closely related to the concept of Hall viscosity, which depends sensitively on geometry and boundary conditions \cite{gromov}. Since the theoretical model employed here addresses only longitudinal transport, we do not make quantitative predictions for Hall measurements. Extending the present analysis to include non-Newtonian effects in the Hall response would require a dedicated theoretical framework and is therefore left for future work.

This work provides compelling evidence for nonlinear magnetotransport in a high-mobility two-dimensional electron gas. We demonstrate that this behavior is driven by a combination of two distinct mechanisms: correlation-driven non-Newtonian viscosity from extended electron collisions and current-induced electron heating. Our analysis successfully separates these contributions, with heating causing a shift in the magnetoresistance peak and nonlinear viscosity enhancing its amplitude. By establishing this dual origin, our study highlights nonlinear conductance measurements as a powerful tool to probe complex, non-equilibrium phenomena in electron fluids, such as pre-turbulent flow and chaotic dynamics. This research not only confirms the existence of a non-Newtonian electron fluid in GaAs quantum wells but also opens a new experimental frontier for investigating quantum hydrodynamics and the potential onset of electronic turbulence in condensed matter systems.

\section {Acknowledgments}
We thank P.S. Alekseev for helpful discussions. This work is supported by FAPESP (São Paulo Research Foundation) Grants No. 2019/16736-2, No. 2021/12470- 8, No. 2024/06755-8, CNPq (National Council for Scientific and Technological Development).

\section{Data Availability}
The data used to generate the figures presented in this study have been deposited in the Zenodo repository and are publicly available under reference \cite{zenodo}

\end{document}